# A High-Precision Multi-Channel TAC and QAC Module for the Neutron Detection Wall *

SHE Qian-Shun (佘乾顺)[1] SU Hong (苏弘) QIAN Yi (千奕) YU Yu-Hong (余玉洪)

Institute of Modern Physics, Chinese Academy of Sciences, Lanzhou 730000, China

**Abstract**

A single width NIM module that includes eight channels TAC (time-to-amplitude converter) and QAC (charge-to-amplitude converter) is introduced in the paper, which is designed for the large neutron wall detector to measure charge (energy) and time interval simultaneously [1]. The module mainly adopts a high precision gated integral circuit to realize TAC and QAC. The input range of TAC is from 30 ns to 1 us, and the input range of QAC is from 40 pC to 600 pC. The linearity error of TAC is lower than 1.28 %, and the time resolution of TAC is less than 0.871 %. The linearity error of QAC is lower than 0.81 %, and the resolution of QAC is better than 0.936 %.

## 1 Introduction

With the national scientific project of Cooling Storage Ring of the Heavy Ion Research Facility (HIRFL-CSR) built in Lanzhou, the External Target Facility (ETF) for HIRFL-CSR is also under construction. The neutron wall is a key detector of the ETF [2], and it is used to measure neutrons with energies from several tens of MeV to 1 GeV.

The neutron wall detector is a large array detector which is composed of 252 detection units, and the time-of-flight method is used to detect neutron. The start detector at the front of the neutron wall produces the start signal of the time-of-flight, and the neutron wall detector produces the stop signal of the time-of-flight. The structure of eight channels TAC and QAC NIM module is shown in Fig.1; it includes TAC circuit, QAC circuit, discriminator and signal splitter.

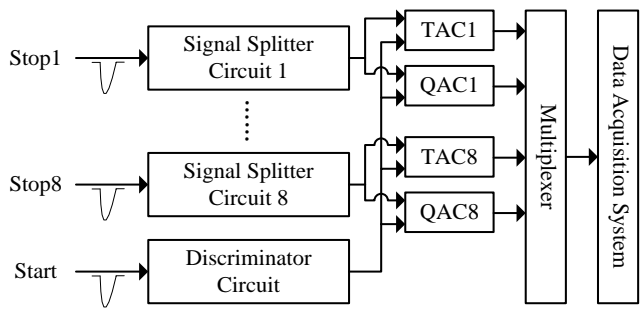

Fig.1. Structural diagram of eight channels TAC and QAC module

* Supported by National Natural Science Foundation of China (11005135 and 11079045) and National Instruments Special of China (2011YQ12009604)
1) E-Mail: sheshun322@impcas.ac.cn

After being discriminated and shaped, the start signal from the start detector triggers the TAC circuit to start integral and the QAC circuit to prepare to do integral. After being split, the stop signal from the neutron wall detector is fed into the TAC and QAC circuit. On the one hand, the stop signal after being discriminated and shaped triggers the TAC circuit to stop integral, so that the time interval between the start and stop signal is converted to a proportional voltage; on the other hand, the stop signal is sent into the QAC circuit, and converted into a proportional voltage on the integral capacitor after current splitter. The voltage amplitude which is formed by the TAC and QAC is converted to a digital signal by data acquisition system after multiplexer.

## 2    The design of TAC

The TAC technology is widely used in the time-of-flight spectrometer at present [3]. Realization of TAC is to convert the time interval into a proportional voltage. After the initial measurement, the time interval, between START signal given by the start time detector at the front of the reaction target and STOP signal which is the stop sign of flight time of neutron given by neutron wall detector, is about 200 ns. So with careful study and comparison [4], a new TAC circuit, a start-stop type TAC, has been proposed which is constructed based on a high precision constant current source and the gated integral. It is characterized by simple circuit structure, high conversion precision , fast discharge, and so on.

The integrating amplifier circuit of TAC unit is mainly composed of a constant current source circuit, gated integral circuit and amplifier circuit, the circuit diagram is shown in Fig.2. The general idea is that we use a gated integral to convert the current from a constant current source into voltage, and the integral time named $T_{GATE}$ is time interval between START signal and STOP signal. The voltage will be maintained for data acquisition system. Then a control signal named $T_{CLEAR}$ discharges the voltage on capacitor C7.

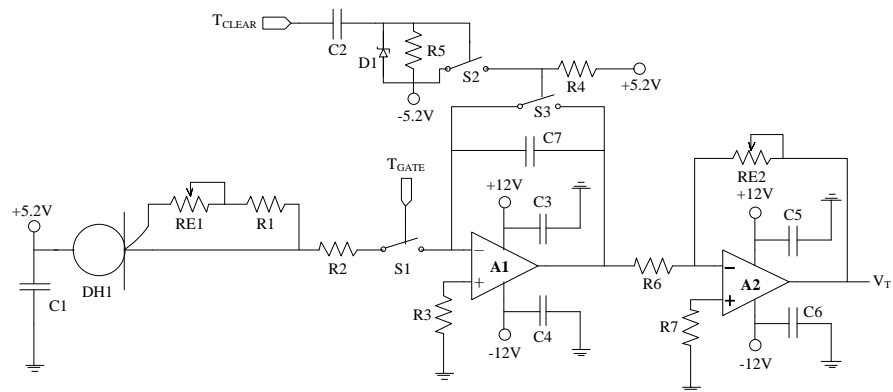

Fig.2. Integral amplifier circuit diagram of TAC

The constant current source circuit consists of C1, DH1, RE1, and R1. DH1 is a constant current source device which has good voltage stability and temperature stability, the current stability (S) is from 0.02 %/V to 0.05 %/V. The temperature coefficient ($\alpha$MAX) is from $1.0 \times 10^{-4}$/℃  to $3 \times 10^{-4}$/℃. This circuit can steadily output current from 0.3 mA to 25 mA by adjusting the potentiometer RE1. In order to prevent the potentiometer from adjusting to zero, it should be connected a small resistor R1 behind RE1.

The high-speed DMOS analog switch S1, S2 and S3, FastFET Op Amps A1 and A2, the integral capacitor C7, three resistors R3, R4 and R5, one diode D1, three capacitors C2, C3, and C4 are used to compose the gated integral circuit with the reset function. The main function of this part is integrating current from the constant current source and maintaining the output voltage. A high-performance mica capacitor C7 is used as the integral capacitor, C2 for the effective isolation, the C3, and C4 for decoupling capacitors.

If one single switch, for example S3, is used to the voltage on the C7 which can't be discharged successfully, especially for negative voltage. In order to discharge the voltage on capacitor C7 completely, the substrate voltage of switch S3 is set to the minus 5.2 V. However, it is not possible to switch the S3 on or off directly by a drive signal from CPLD, if the substrate voltage of switch is set to the minus 5.2 V.

A driven type DMOS switch circuit is designed that is composed of two switches S2 and S3, two resistors R4 and R5, and one diode D1. The control signal from CPLD drives the switch S2 on or off, which can make the control voltage to switch S3 changing from positive 5.2 V to negative 5.2 V, which is the control voltage to S3 is 5.2 V when the S2 is switched off, otherwise the control voltage to S3 is minus 5.2 V when the S2 is switched on. This circuit can ensure switch 3 to cut-off or turn-on completely, so that the gated integral circuit can work properly. The inverting amplifier consists of operation amplifier A2, resistors R6, R7 and RE2, the gain can be changed by adjusting the RE2.

A front-edge timing discriminator is designed to discriminate the signal from the detector for time analysis. The front-edge timing circuit of the START signal is shown in Fig.3. The circuit includes a new rapid discriminate device MA1. The START signal is firstly fed into the rapid isolation circuit, and then enters into the front-edge timing circuit in order to ensure the consistency of the signal delay. After test, the propagation delay of signal from input to output of the entire front-edge timing circuit is less than 1 ns, the time walk range is about 4 ps and 6 ps in theory. This circuit has greatly improved the precision of TAC.

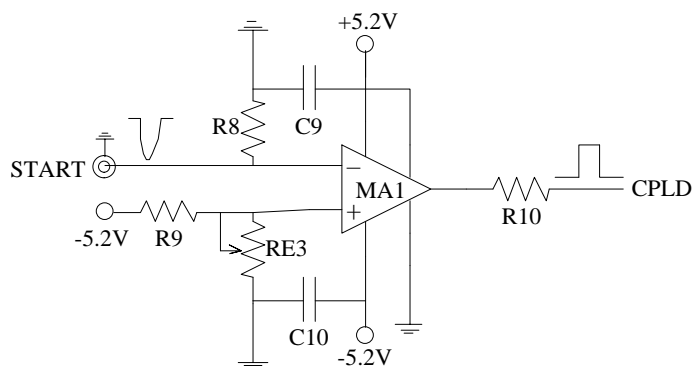

Fig.3. Front-edge timing circuit of START

The integral time ($T_{GATE}$) in the TAC circuit is determined by the input signals START and STOP. The duration between START and STOP is the amount of time to be converted. The START and STOP signals are converted to TTL signal by front-edge timing circuit and then sent into CPLD, the CPLD produce $T_{GATE}$ signal and $T_{CLEAR}$ signal. The falling-edge of $T_{RANG}$ generates the control signal of multiplexer and the sampling clock signal SAMP$_{CLK}$. The SAMP$_{CLK}$ controls correspondingly the data acquisition card to read-out the voltage signals of eight channels TAC and eight channels QAC. Fig.4 shows the timing chart of the TAC operation.

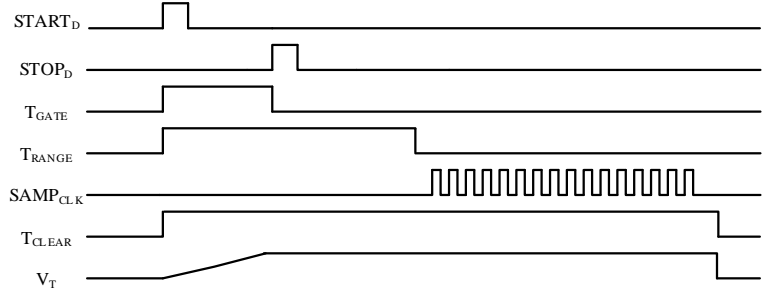

Fig.4. Timing chart of TAC operation

START$_D$ and STOP$_D$ signals, which are from START and STOP, and formed by the front-edge timing circuit, corresponds to the integral start time and stop time. START$_D$ and STOP$_D$ in CPLD are formed into switch control signal (T$_{GATE}$) of the integral, and its width is the time interval to be measured. The charged voltage on integral capacitor C7 is given by Eq.(1):

$$V_T = \frac{I*(T_{STOP} - T_{START})}{C7} = \frac{I*T_{GATE}}{C7} \qquad (1)$$

T$_{RANGE}$ is triggered by the START$_D$ and set as the maximal input range of TAC to be measured. SAMP$_{CLK}$ which is triggered by T$_{RANGE}$ and sent to the data acquisition system is sample clock for TAC and QAC module with eight channels. T$_{CLEAR}$ is triggered by the START$_D$ and used as clear signal, when it is at high means that the switch is off.

## 3    The design of QAC

The output signal from the neutron wall detector is very fast. If we use a charge sensitive preamplifier to process the signal, it will cause very slow trailing edge. So it is not good to accumulate the signal when the count rate of input signal is high. Therefore we propose a QAC approach to realize rapidly and accurately QAC based on reference [5-6]. In the process of research and development, we continue to improve and optimize the design of the QAC circuit, and make the circuit can accurately handle the fast current signal, and its prominent characteristic is the fast convert speed, large range of input signal (40 pC - 600 pC), high precision, low power, high integral, simple circuit structure, operation stability. QAC circuit mainly consists of a current splitter circuit, gated integral circuit and amplifier. Gated integral circuit and amplifier are the same as one in TAC unit, the design of QAC circuit is shown in Fig.5.

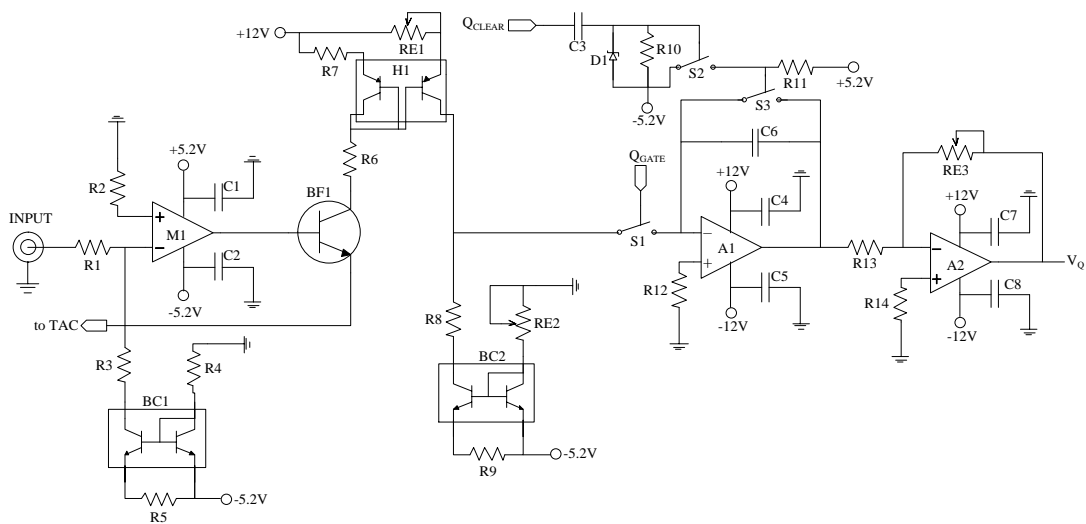

Fig.5. Circuit diagram of QAC

The input charge (current) enters into the gated integral circuit after the rapid current splitter circuit. It can be charged on the integral capacitor by controlling the analog switch. It can get voltage amplitude proportional to the input signal on the integral capacitor. After sampled, the voltage amplitude will be discharged by the discharge switch, so that the potential at each end of the capacitor return to initial state.

M1, BC1, BC2, BF1, H1 are composed of the fast current splitter circuit, so we choose a wide band amplifier M1 and the transistor BF1  with cutoff frequency up to 3 GHz, which can improve the sensitivity of the input circuit as well as the signal-to-noise ratio, and achieve impedance transformation. The micro-current source consists of R3, R4, R5, and BC1, which can realize precise current adjustment. It can not only compensate offset current on negative input of the amplifier M1, but also ensure the input is zero potential and zero current in cooperation with the depth of the negative feedback circuit when no signal input. Another current source consists of R8, R9, RE2and BC2, which ensure that the input of the gated integral is zero current by adjusting RE2 when no signal input, and the output pedestal voltage step is completely eliminated. The proportional current source consists of R7, RE1, and H1. As long as changing the proportional relationship of the RE1 and R7, we can change the ratio of the input current and the current fed to the gated integral. In this way, it can initially amplify the small current signal and improve signal-to-noise ratio. If the input signal is a big current signal, it can be attenuated down by this proportional current source. So this fast current splitter circuit can play a role of attenuation network, which is beneficial to simplify the circuit structure, and can greatly improve the measurement accuracy and measurement range.

Fig.6 shows the timing chart of the QAC operation. $START_D$ is a common signal for TAC and QAC, the STOP signal is the output of the detector, $Q_{GATE}$ signal is triggered by the $START_D$, the width of $Q_{GATE}$ can be set in the program according to the actual condition, but which must include the pulse duration of the STOP signal to ensure that STOP was completely integrated. $Q_{CLEAR}$ is also triggered by $START_D$, when all output signals of eight channels TAC and QAC have been sampled, the $Q_{CLEAR}$ change into a low-level to discharge the voltage on integral capacitor.

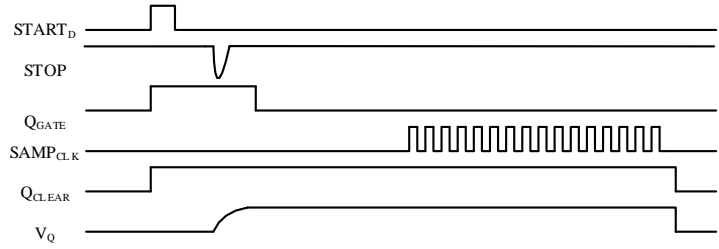

Fig.6. Timing chart of QAC operation

When the leading-edge of $Q_{GATE}$ signal is coming, the normally open switch S1 is closed, the normally-close switch S2 is opened, the system begins to work, and the STOP signal is integrated. When the falling-edge of the $Q_{GATE}$ signal is coming, the constant-opened switch S1is opened, DAQ system begins to acquire data from the whole channels of this system, and the state of the normally-close switch S2 is unchanged at the same time. When all signals of the channels have been acquired, the normally close switch S2 is closed, the integral capacitor discharges charge, then this system begins to prepare for the next data acquisition.

## 4    Performance

In order to test the function of this NIM module, we set up a test system in the experimental field, the structural diagram of functional test system is shown in Fig.7. The system has been tested with cosmic ray for a period of 30 days with non-stop work. In this system, a NI PXI-6133 card is employed as a data acquisition card.

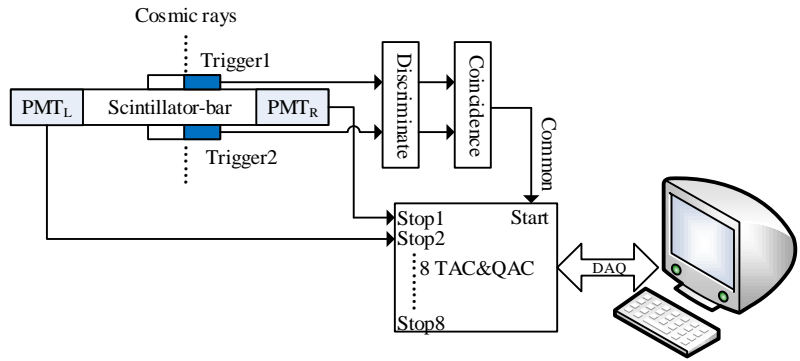

Fig.7. Structural diagram of functional test

The TAC and the QAC have a common start and work at the same time. The result is shown in Figs.8 and 9. The symbol 'Entries' is the count of events, the symbol 'Mean' is the average voltage, the symbol 'RMS' is the standard deviation of the amplitude, the test results with the Cosmic ray of TAC and QAC conform to the Gaussian distribution.

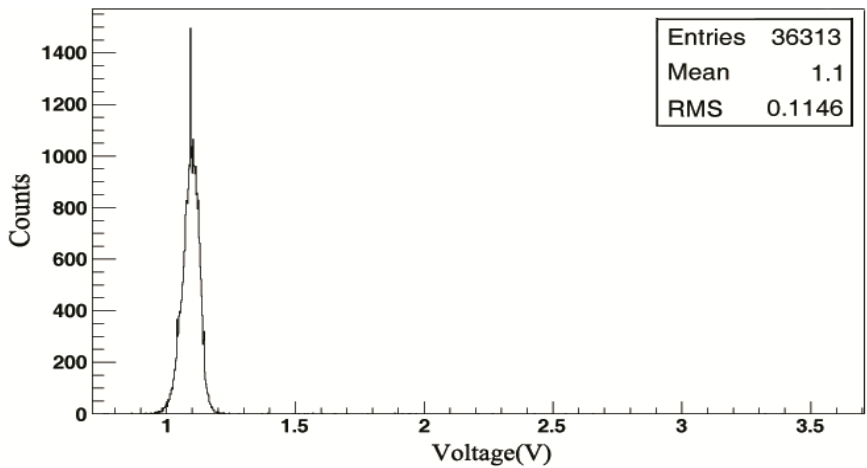

Fig.8. Functional test result of TAC

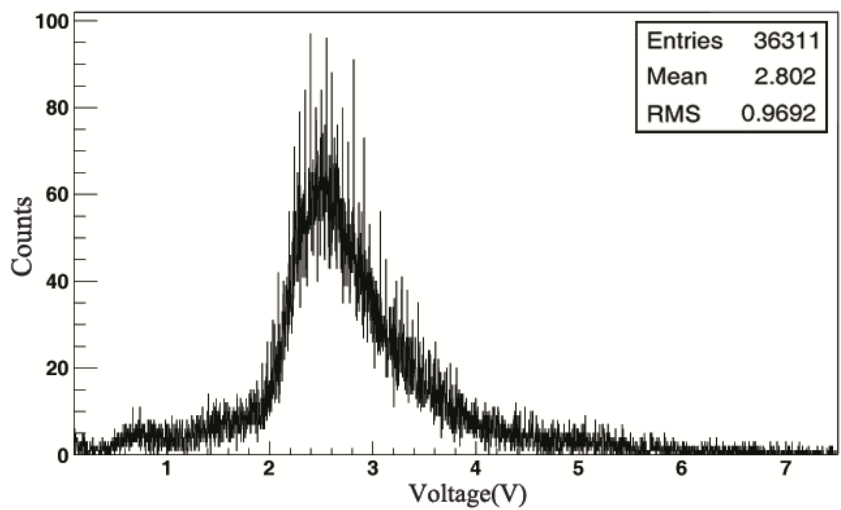

Fig.9. Functional test result of QAC

The performance of this NIM module was tested in the laboratory. Fig.10 shows the structural diagram of performance test system. The Phillips CAMAC Model 7120 was selected as the signal source, a NI Corporation DAQ card PXI-6133 and an industrial computer were used to compose the data acquisition system.

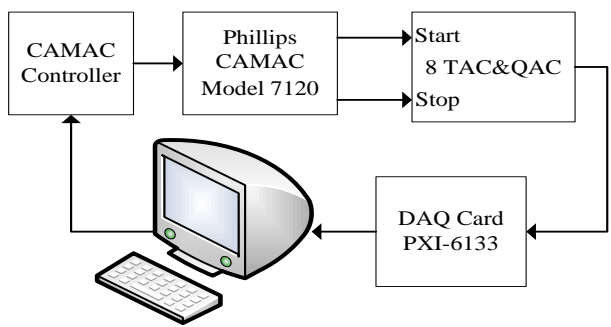

Fig.10. Structural diagram of performance test

Parameters of the TAC and the QAC are shown in Tables 1 and 2, each test point was tested more than 10 million events. The transfer curve and linearity errors of TAC are shown in Fig.11, The longer the time interval under tested, the better the resolution. The transfer curve and linearity errors of QAC are shown in Fig.12, The more the input charge, the better the resolution.

**Table 1**

Parameters of the TAC

| Measurement range | 30 nsec to 1 usec |
|---|---|
| Holding capacitor | 1.33 nF |
| Full scale differential output | 10 V ($\pm$5V) |
| Discharge time | <200 nsec |
| Output voltage noise | <4.6 mV (rms) |
| Output offset voltage | <$\pm$4.1 mV |
| Linearity error | 0.051 % to 1.280 % |
| Time resolution | 0.066 % to 0.871 % |
| Absolute resolution | 261 ps to 656 ps |

**Table 2**

Parameters of the QAC

| Measurement range | 40 pC to 600 pC |
|---|---|
| Holding capacitor | 330 pF |
| Full scale differential output | 10 V ($\pm$5V) |
| Discharge time | <200 nsec |
| Output voltage noise | <5.2 mV (rms) |
| Output offset voltage | <$\pm$4.7 mV |
| Linearity error | 0.003 % to 0.810 % |
| Energy resolution | 0.043 % to 0.936 % |
| Absolute resolution | 240 fC to 374 fC |

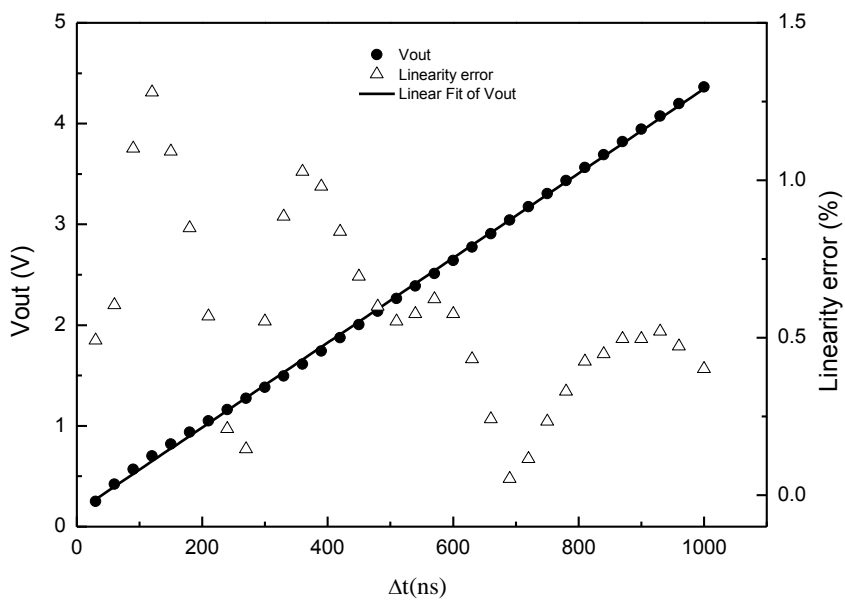

Fig.11. Transfer curve and linearity errors of TAC

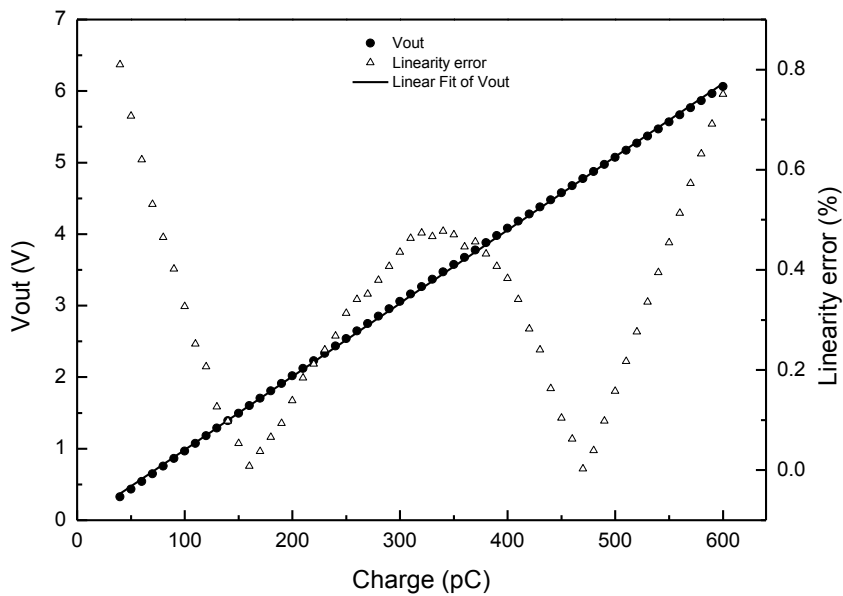

Fig.12. Transfer curve and linearity errors of QAC

## 5    Summary

This NIM module and a normal DAQ system was used to constitute a TDC and QDC system, a comparison test has been implemented, When chose Phillips CAMAC Model 7120 as the signal source, the related performance of this system is superior to some commercial TDC and QDC modules, the linearity error and resolution of this NIM module are better than the commercial CAMAC module about 0.1 %. The TAC and QAC circuit developed by us has the characteristics of high processing speed, simple circuit structure, high-precision and low power dissipation, but also has a low manufacture cost. In particular, this NIM module can be readout the time interval and charge information simultaneously with a same signal. The resolution and linearity error are litter worse at the very low measurement range, which is needed to be improved. This module can be used widely to construct front-end read-out electronics system for large array detector.